\documentclass{aa}
\usepackage{epsfig}

\begin{document}

\title{Non--axisymmetric accretion on the classical TTS RW~Aur~A\thanks{Based
      on observations made with the Nordic Optical
      Telescope, operated on the island of La Palma jointly by Denmark,
      Finland, Iceland, Norway, and Sweden, in the Spanish Observatorio del
      Roque de los Muchachos of the Instituto de Astrofisica de Canarias.}}

\author{P.P.~Petrov\inst{1}\thanks{on leave from the Crimean
 Astrophysical Observatory}, G.F.~Gahm\inst{2}, J.F.~Gameiro\inst{3,4},
                             R.~Duemmler\inst{1}, I.V.~Ilyin\inst{1},
                             T.~Laakkonen\inst{1},
                             M.T.V.T.~Lago\inst{3,4}, I.~Tuominen\inst{1}}

\offprints{P.P.\ Petrov; Peter.Petrov@Oulu.Fi}
%\email{Peter.Petrov@Oulu.Fi}

\institute{Astronomy Division, P.O.~Box 3000, FIN--90014 University of
Oulu, Finland
\and Stockholm Observatory, SE--133 36 Saltsj\"{o}baden, Sweden
\and CAUP, Universidade do Porto, Rua das Estrelas, PT--4150 Porto, Portugal
\and Departamento de Matem\'atica Aplicada, Rua das Taipas, 135, PT--4150,
Portugal}

\date{Received date; Accepted date}

\authorrunning{P.P.\ Petrov et al.}
\titlerunning{Non--axisymmetric accretion on RW~Aur~A}

\abstract{
  High--resolution spectroscopic monitoring of the exceptionally
  active classical T~Tauri star (CTTS) RW~Aur~A was carried out in
  three seasons of 1996, 1998 and 1999 with simultaneous B, V
  photometry. The high quality spectra revealed a multicomponent
  structure of the spectrum, which includes: 1) a veiled photospheric
  spectrum of a K1--K4 star, 2) broad emission lines of neutrals and
  ions, 3) narrow emission lines of \ion{He}{i}
  and \ion{He}{ii}, 4) red--shifted
  accretion features of many lines, 5) shell lines at about the
  stellar velocity, 6) blue--shifted wind features and 7) forbidden
  lines.\\
  Periodic modulations in many spectral features were found.  The
  photospheric absorption lines show sinusoidal variations in radial
  velocity with an amplitude of $\pm6$\,km\,s$^{-1}$ and a period of
  about $2\fd77$. The radial velocities of the narrow emission lines
  of He vary with the same period but in anti--phase to the
  photospheric lines. The equivalent widths of the narrow emissions
  vary with a phase--shift with respect to the velocity curve. The
  strength of the red--shifted accretion components of Na~D and other
  lines is also modulated with the same period. The broad emission
  lines of metals vary mostly with the double period of about $5\fd5$.\\
  One unexpected result is that {\it no correlation} was found
  between the veiling and the brightness, although both parameters varied
  in wide ranges.  This is partly due to a contribution of the shell
  absorption to the photospheric line profiles, which make them vary
  in width and depth thus simulating lower veiling.\\
  The spectral lines of the accreting gas show two distinct
  components: one is formed at low velocity at the beginning of the
  accretion column, and the other at high velocity near the stellar
  surface. The low velocity components are strong in low excitation
  lines of neutrals, while the high velocity components are strong in
  high excitation lines of ions, thus showing the gradients of
  temperature and density along the accretion column.\\
  Most of the observed features can be interpreted in the framework of
  non--axisymmetric magnetospheric accretion, but the origin of this
  asymmetry can be explained in different ways. We consider two
  possible models. The first model suggests that RW~Aur~A is a binary
  with a brown dwarf secondary in a nearly circular orbit with a
  period of $2\fd77$. The orbiting secondary generates a moving
  stream of enhanced accretion from one side of the disk towards the
  primary.  The other model assumes that RW~Aur~A is a single star
  with a rotational period of $5\fd5$ and with two footpoints of
  channeled accretion streams within a global magnetosphere which is
  tilted relative to the rotational axis or otherwise
  non--axisymmetric. Both models can explain qualitatively and
  quantitatively most of the observed variations, but there are
  some details which are less well accounted for.
}
\maketitle

\section{Introduction}
\label{Intro}

RW~Aurigae is one of the brightest T Tauri stars (TTSs).
It is also a most unusual and extreme object of its kind, and has always
played a joker in the exploration of the nature of young low mass
pre--main sequence objects.

RW~Aur is a resolved triple star system (Ghez et al.\ \cite{Ghez93}),
and has a bipolar jet rooted at the brightest component, RW~Aur~A
(e.g.\ Mundt \& Eisl\"{o}ffel \cite{Mundt98},
Dougados et al.\ \cite{Dougados00b}). Strong and rapid, but irregular
brightness variations are common --- usually measured for component A, which
dominates the flux (see e.g.\ Gahm et al.\ \cite{Gahm93}; Herbst et al.\
\cite{Herbst94}), and the star is a bright infrared source, with a
rather flat spectrum longward of 1 $\mu$m (Cohen \& Schwartz
\cite{Cohen76}; Ghez et al.\ \cite{Ghez97}).

Already Joy (\cite{Joy45}) commented on the magnificent spectrum with strong
and broad emission lines. Weak, presumably photospheric, absorption
lines indicate a star of spectral type K (e.g.\ Mundt \& Giampapa
\cite{Mundt82}, Valenti et al.\  \cite{Valenti93}), and
Hartmann et al.\ (\cite{Hart86}) found that these lines vary in radial
velocity.
The weakness of the lines indicates the presence of continuous
and/or line excess emission --- the spectrum is veiled.

The emission line spectrum is very complex, and many lines are
blended with each other. Some of the broad emission
lines also have narrow emission components, and some have
absorption components: broad red--shifted, central and/or slightly
blue--shifted in velocity. All these components have been found to
undergo strong and rapid (hours/days)
changes in fluxes, profiles, and radial velocities (Gahm \cite{Gahm70};
Appenzeller \& Wolf \cite{Appen82}; Hartmann \cite{Hart82}; Mundt \&
Giampapa \cite{Mundt82}; Appenzeller et al.\ \cite{Appen83}; Grinin et
al. \cite{Grinin83}; Stout--Batalha \& Batalha \cite{Stout00}).

Many ideas on the cause of the photometric and spectral variability
observed on classical TTSs, like RW~Aur~A, have seen light over the
years. Presently, the concept of magnetospheric accretion has
been successful in explaining many observed phenomena
(see Hartmann \cite{Hart98}, and references therein).
Irregular accretion to heated
regions at the stellar poles has been recognized as a process for
producing irregular light variability (variable veiling hot
spots), while gas blobs flowing through the magnetosphere
could cause variability of the emission line spectrum (see e.g. Gullbring
et al. \cite{Gull96} and references therein).

The magnetospheric accretion model predicts a correlation between
veiling and brightness. Such correlations have been reported (e.g.\
Gahm et al.\ \cite{Gahm95}; Hessman \& Guenther \cite{Hess97}; Chelli
et al.
\cite{Chelli99}), but rather little systematic, long--term
spectroscopic and photometric monitoring has been made in order to
explore this effect.

The original aim of the present program was to check this particular
prediction. One of the selected objects was RW~Aur~A, which shows
clear evidence of variable accretion in several spectral lines along
with the chaotic photometric variability. Surprisingly, we found
practically {\it no} correlation between the degree of veiling and the
brightness, as will be discussed below. Instead, we discovered
periodic radial velocity changes in the photospheric lines, and also
periodic phenomena in practically all other spectral components (Gahm
et al.\ \cite{Gahm99}, hereafter called Paper~I).

In Paper~I, we focussed on the possibility that RW~Aur~A is a close
binary with a secondary of very low mass, possibly a brown dwarf,
moving in a nearly circular orbit at less than 10 solar radii from the
primary.  We reserved a full discussion of all spectral features, and
their complex and interwoven variabilities, for a more extensive
presentation. In addition, we have also collected a new series of
observations of the star at the Nordic Optical Telescope in 1999.  The
present paper summarizes our findings.

\section{Observations}
\label{Obs}

\begin{table*}
\caption{The first 3 columns constitute the observing log. SNR is the
	   signal--to--noise ratio in the continuum near H$\alpha$. The next
       4 columns present the radial velocities for the weak absorption
       lines (WAL) and the \ion{He}{i} narrow emission line together with 
       their standard deviations ($\sigma$); 
       all velocities are in the stellar restframe,
       i.e.\ the mean stellar radial velocity of +16\,km\,s$^{-1}$ has been
       subtracted. Columns 9 and 10 give the veiling factor and its standard
       deviation. The last 3 columns give the simultaneous UBV
       photometry.}
\label{Tab1}
\begin{center}
\begin{tabular}{cccr|rlrl|cc|rrr}
\hline
{Date}  &     HJD 245... & file  &  SNR &   WAL & $\sigma_{\rm WAL}$ 
        & \ion{He}{i} & $\sigma_{\rm \ion{He}{i}}$ & veil 
        & $\sigma_{\rm veil}$  &  V  &  B--V & U--B\\
\hline
03/04  Dec  95 & 0055.521 & 06249 &  120 &  +4.8 & 1.6 &  +3 & 1.0 &   3.9
& 0.7 &   --- &  --- &  --- \\
05/06  Dec  95 & 0057.565 & 06424 &  120 &  +8.8 & 2.8 &  +7 & 2.0 &   3.1
& 0.5 &   --- &  --- &  --- \\
\hline
24/25  Oct  96 & 0381.586 & 07032 &  140 &  +3.5 & 1.4 & +11 & 3.0 &   4.6
& 0.7 &  9.82 & 0.55 &  --- \\
24/25  Oct  96 & 0381.631 & 07036 &  130 &  +0.8 & 1.4 & +15 & 2.0 &   3.4
& 0.4 &  9.94 & 0.52 &  --- \\
25/26  Oct  96 & 0382.570 & 07097 &  200 & --1.0 & 2.0 &   0 & 2.0 &   7.0
& 2.0 &  9.98 & 0.70 &  --- \\
25/26  Oct  96 & 0382.624 & 07103 &  150 & --2.6 & 2.0 &  +3 & 2.0 &  10.0
& 2.0 & 10.04 & 0.68 &  --- \\
26/27  Oct  96 & 0383.536 & 07179 &  150 & --7.6 & 1.5 & +17 & 2.0 &   2.6
& 0.3 & 10.13 & 0.66 &  --- \\
26/27  Oct  96 & 0383.583 & 07183 &  180 & --9.4 & 1.5 & +17 & 1.0 &   2.8
& 0.3 & 10.11 & 0.65 &  --- \\
27/28  Oct  96 & 0384.568 & 07255 &  150 &  +2.5 & 1.5 & --3 & 1.0 &   1.5
& 0.2 & 10.02 & 0.61 &  --- \\
27/28 Oct 96 & 0384.617 & 07259 &  150 &  +3.6 & 1.5 & --5 & 1.0 &   1.7 &
0.2 & 10.16 & 0.62 &  --- \\
28/29 Oct  96 & 0385.580 & 07323 &  220 &  +6.4 & 2.5 & --2 & 3.0 &   2.5 &
0.3 & 10.04 & 0.63 &  --- \\
28/29  Oct  96 & 0385.652 & 07329 &  180 &  +6.0 & 2.5 & --4 & 4.0 &   3.5
& 0.6 & 10.16 & 0.61 &  --- \\
29/30  Oct  96 & 0386.576 & 07388 &  200 &  +0.1 & 2.1 & +26 & 2.0 &   1.5
& 0.1 &  9.95 & 0.45 &  --- \\
29/30  Oct  96 & 0386.638 & 07392 &  170 & --1.2 & 2.1 & +28 & 2.0 &   1.2
& 0.2 &  9.96 & 0.46 &  --- \\
30/31  Oct  96 & 0387.721 & 07455 &  100 &  +9.1 & 3.0 &  +5 & 3.0 &   2.7
& 0.6 &  9.88 & 0.52 &  --- \\
31/01  Nov  96 & 0388.511 & 07491 &  230 & --3.9 & 3.0 &  +8 & 4.0 &   1.3
& 0.2 & 10.04 & 0.73 &  --- \\
31/01  Nov  96 & 0388.573 & 07495 &  200 & --5.3 & 3.0 & +13 & 3.0 &   1.7
& 0.2 &  9.99 & 0.75 &  --- \\
\hline
17/18  Aug  97 & 0678.743 & 09972 &   52 &  +9.0 & 3.0 &  +1 & 1.0 &   5.2
& 0.8 &  ---  & ---  &  --- \\
19/20  Aug  97 & 0680.719 & 10168 &   40 &   0.0 & 2.0 &  -- &  -- &   1.1
& 0.2 &  ---  & ---  &  --- \\
19/21  Aug  97 & 0681.713 & 10258 &   50 &   0.0 & 3.0 &  +14 & 1.0 &   2.3
& 0.4 &  ---  & ---  &  --- \\
21/22  Aug  97 & 0682.719 & 10344 &   40 &  -1.0 & 2.0 &  -- &  -- &   0.7
& 0.1 &  ---  & ---  &  --- \\
\hline
04/05  Nov  98 & 1122.554 & 16418 &   60 &  +7.0 & 2.0 &  +6 & 3.0 &   2.6
& 0.4 & 10.69 & ---  &  --- \\
04/05  Nov  98 & 1122.741 & 16448 &   80 &  +5.0 & 2.0 & +10 & 3.0 &   2.6
& 0.5 &  ---  & ---  &  --- \\
05/06  Nov  98 & 1123.545 & 16652 &   70 & --6.3 & 2.1 & +18 & 5.0 &   1.6
& 0.2 & 10.67 & 0.57 &  --0.44 \\
05/06  Nov  98 & 1123.622 & 16666 &   60 & --6.3 & 2.1 & +19 & 5.0 &   1.0
& 0.1 &  ---  & ---  &  --- \\
06/07  Nov  98 & 1124.543 & 16756 &  100 &  +3.5 & 2.6 & --13 & 3.0 &   1.0
& 0.1 & 10.58 & 0.60 & --0.48 \\
06/07  Nov  98 & 1124.725 & 16784 &  100 &  +8.2 & 2.6 & --12 & 2.0 &   1.6
& 0.1 &  ---  & ---  &  --- \\
07/08  Nov  98 & 1125.545 & 16907 &   80 &  +0.8 & 3.0 & +12 & 1.5 &   3.6
& 0.5 & 10.66 & 0.71 &  --0.10 \\
08/09  Nov  98 & 1126.657 & 17060 &   55 & --6.7 & 3.0 & +10 & 5.0 &   2.6
& 0.2 & 10.44 & 0.70 &  --0.16 \\
08/09  Nov  98 & 1126.712 & 17076 &   40 & --9.2 & 3.0 & +11 & 3.0 &   2.3
& 0.2 &  ---  & ---  &  --- \\
09/10  Nov  98 & 1127.546 & 17161 &   50 &  +2.0 & 2.0 & --3 & 3.0 &   1.4
& 0.2 & 10.17 & 0.76 &  --- \\
09/10  Nov  98 & 1127.718 & 17189 &   60 &  +6.9 & 3.0 & --3 & 2.0 &   2.7
& 0.3 &  ---  & ---  &  --- \\
\hline
19/20  Oct  99 & 1471.754 & 22344 &   80 &  +6.4 & 1.4 &  +8 & 1.0 &   3.9
& 0.8 & 10.29 & 0.76 &  --- \\
20/21  Oct  99 & 1472.758 & 22506 &   70 & --9.9 & 1.2 & +24 & 2.0 &   2.7
& 0.4 & 10.45 & 0.67 &  --- \\
21/22  Oct  99 & 1473.723 & 22649 &   60 &  +2.9 & 2.2 &  +5 & 2.0 &   8.0
& 2.0 & 10.63 & 0.67 &  --- \\
23/24  Oct  99 & 1475.753 & 22838 &   50 & --0.8 & 2.6 &  +9 & 5.0 &   1.5
& 0.2 & 10.98 & 0.64 &  --- \\
25/27  Nov  99 & 1509.633 & 23018 &   60 & --5.5 & 1.6 & +20 & 2.0 &   1.1
& 0.2 & 10.50 & 0.62 &  --- \\
27/28  Nov 99 & 1510.709 & 23212 &  120 &  +2.4 & 1.2 &  +8 & 1.0 &   2.6 &
0.3 & 10.20 & 0.63 &  --- \\
\hline
\end{tabular}
\end{center}
\end{table*}

\begin{figure}
\centerline{\resizebox{5cm}{!}{\includegraphics{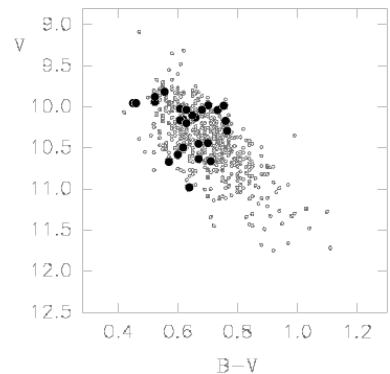}}}
%%%%%%%width of the figure=5cm; {!} keeps the proportions like in the original
\caption{Colour--magnitude diagram. Open circles: data from the Herbst
catalogue; filled circles: our observations.}
\label{v_bv.ps}
\end{figure}
\begin{figure}
  \resizebox{\hsize}{!}{\includegraphics{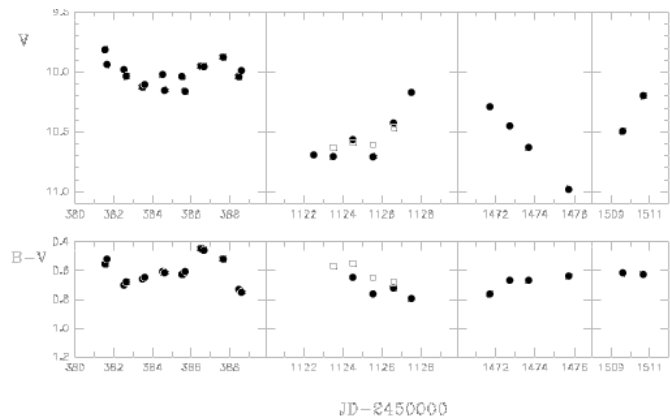}}
  \caption{Light and colour curves for the seasons 1996, 1998, and 1999.
    Filled circles: CCD observations at NOT; open squares: photoelectric
    observations at the Swedish telescope.}
  \label{v.ps}
\end{figure}

High--resolution spectra were collected with the SOFIN \'echelle
spectrograph (Tuominen et al.\ \cite{Tuo99}) at the 2.56m Nordic
Optical Telescope (NOT) during several observing periods in 1995,
1996, 1997, 1998 and 1999 (Table~\ref{Tab1}).  We used the 3rd camera, which
provides a spectral resolution of about 12\,km\,s$^{-1}$ with the
entrance slit of $1\farcs7$ ($R=26\,000$).  The secondary components
(RW~Aur~B and C) were mostly outside of the entrance slit, but when
the seeing was not good enough, the secondary could contribute to the
observed spectrum.  The flux ratio of the components A/BC is 18 at
5550\,\AA\ and 12 at 6750\,\AA\ (Ghez et al.\ \cite{Ghez97}). From this,
we can estimate the contribution from the secondary stars as a few percent
(in the red continuum) in the worst case.  In the night of best seeing
we took one spectrum of RW~Aur~BC separately.  The only strong
emission line is H$\alpha$, and the lines of \ion{He}{i} and \ion{Ca}{ii}
are weak
and narrow. All these lines are very much stronger in the spectrum of
RW~Aur~A. We conclude that the possible contribution from RW~Aur~BC in
the emission lines was less than 1 percent.

The 43 spectral orders registered in one CCD frame cover a region of 3500 to
11000\,\AA\ with a length of 140\,\AA\ per order near H$\alpha$. Usually, we took 
two exposures of 20 minutes which were added into one spectrum; this gave a
useful range of 3900 to 9000\,\AA\ with some gaps in the red. In 1996,
we took spectra with a shift in the \'echelle position, so that the whole
spectral range was covered without gaps.
In some nights, two spectra were taken with an interval of a few
hours.

The CCD images of the \'echelle spectra were
obtained and reduced with the 4A software package (Ilyin \cite{Ilyin00}).
The standard procedure involves bias subtraction, estimation of the
variances of the pixel intensities, correction for the master flat field,
scattered light subtraction with the aid of 2D--smoothing splines,
definition of the spectral orders, and weighted integration of the
intensity together with elimination of cosmic spikes.

The wavelength calibration was done with a ThAr comparison spectrum;
one was taken before and one after each individual object exposure to
eliminate any temporal changes in the spectrograph during the exposure. The
wavelength solution incorporates the Gaussian--centred positions,
wavelengths, and times of all detected spectral lines from the two comparison
images. The
wavelength for every pixel in the stellar spectrum is calculated for the
time of
its mid-exposure. The wavelength solution also incorporates the positions of
all detected telluric lines in the stellar spectrum which eliminates the
wavelength shifts caused by the slit effect of the spectrograph.
A typical error of the wavelength scale
in the centre of the image is about 50--100\,m\,s$^{-1}$.
The correction of the spectra for the vignetting function and for the
Earth orbital motion constitute the next steps of the data reduction.
Finally, the continuum was determined by fitting a smooth curve to the ratio of
each individual and the mean spectrum.
The spectra are transformed into the stellar restframe, i.e.\ the average
radial velocity of +16 km\,sec$^{-1}$ is subtracted; all velocities given in
this paper are in the stellar restframe.

Photometric observations were carried out with the stand--by CCD camera
at NOT in B and V. In 1996, CCD exposures were taken
before and after each spectral exposures. In 1998 and 1999,
the CCD photometry was done only before the spectral exposures.
In addition, in 1998, UBV photoelectric photometry
was carried out with the Swedish 0.6m telescope on La Palma.
A few spectra taken occasionally in 1995 and 1997 were not accompanied
by photometry.

The log of the observations is given in Table~\ref{Tab1} together with some
parameters of the spectra discussed in the following sections.

The colour--magnitude diagram in Fig.~\ref{v_bv.ps}, including the
data published by Herbst et al.\ (\cite{Herbst94}), covers over 30 years of
observations.  It shows that during our observations RW~Aur~A varied
in brightness within one magnitude in V.  The light--curves are shown
in Fig.~\ref{v.ps}.

\section{Multicomponent structure of the spectrum}
\label{Multicomp}

\begin{figure}
  \centerline{\resizebox{8cm}{!}{\includegraphics{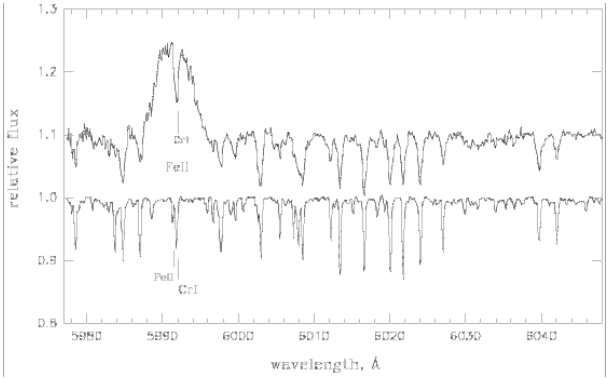}}}
\caption{Upper: the spectrum of RW~Aur~A (average of 1998).
  Note that the broad emission belongs to \ion{Fe}{ii}~5991.37\,\AA, and the
  superimposed narrow absorption to \ion{Cr}{i}~5991.86\,\AA. Lower:
  the spectrum of $\gamma$~Cep (K1\,III--IV), artificially veiled by a
  factor of 3.}
\label{6020.ps}
\end{figure}

\begin{figure}
  \centerline{\resizebox{7cm}{!}{\includegraphics{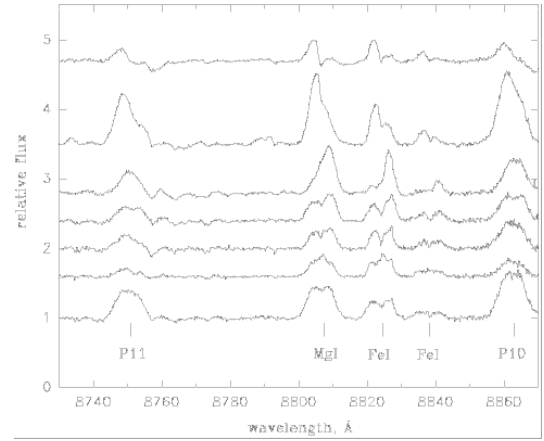}}}
  \caption{An example of night--to--night variability of the broad emission
lines
    in 1998.}
\label{26all.ps}
\end{figure}

\begin{figure}
\centerline{\resizebox{7.5cm}{!}{\includegraphics{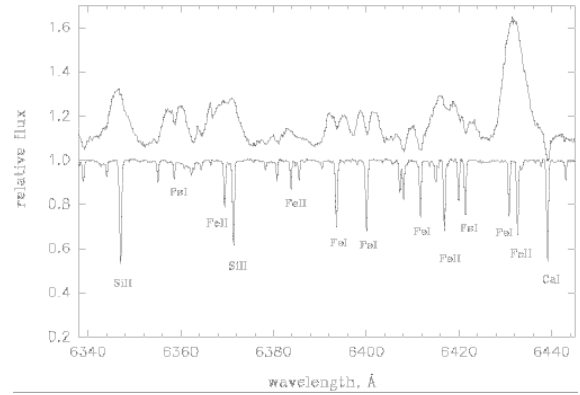}}}
\caption{A fragment of the average spectrum of RW~Aur~A (over all seasons) with
broad emission
lines compared to the spectrum of 41~Cyg (F5\,II). Note that the \ion{Fe}{i}
emissions are split by narrow absorptions,
while the \ion{Fe}{ii}, \ion{Si}{ii} emissions are not.}
\label{35avr.ps}
\end{figure}

\begin{figure}
\resizebox{\hsize}{!}{\includegraphics{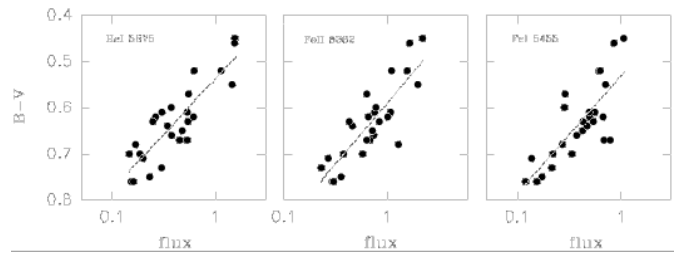}}
\caption{Fluxes of the broad emission lines correlate with colour (fluxes
       in units of $10^{-12}$\,erg\,cm$^{-2}$\,\AA$^{-1}$).}
\label{bv_flux.ps}
\end{figure}

In this section, we give a brief overview of different spectral
features. Several components can be distinguished in the spectrum.
The {\it weak absorption lines} (WALs) can be identified as the highly veiled
photospheric spectrum of a K1--K4\,V star (average veiling = 3; see 
Fig.~\ref{6020.ps}).
In the spectral classification, we try to avoid the lines originating from
low excitation levels, because they can be enhanced by additional
absorption in the accreting gas well above the photospheric level
(Stout--Batalha et al.\ \cite{Stout00}). This effect is quite strong
in density sensitive lines, like \ion{Ba}{ii}~6141\,\AA\ 
($\chi_{\rm exc}=0.7$\,eV).
More details about
the accretion enhancement of absorption lines, {or \it shell lines},
are given in Sect.~\ref{specaccret}.
The width of the WALs, if interpreted only as rotationally
broadened, corresponds on average to $v\,\sin i=30$\,km\,s$^{-1}$,
but was found variable from night to night from 16 to 40\,km\,s$^{-1}$.
Variability in radial velocity of the WALs is discussed in
Sect.~\ref{periods}.
Hereafter, we refer to these weak absorption lines
as the ``photospheric spectrum'', although there may be a contribution
from the layers above the photosphere (the shell).
The spectral region, where the photospheric
spectrum is least blended with emission lines, is shown in Fig.~\ref{6020.ps}.
In some nights, the veiling was so high that very little of
the photospheric spectrum remained visible.

\begin{figure}
\resizebox{\hsize}{!}{\includegraphics{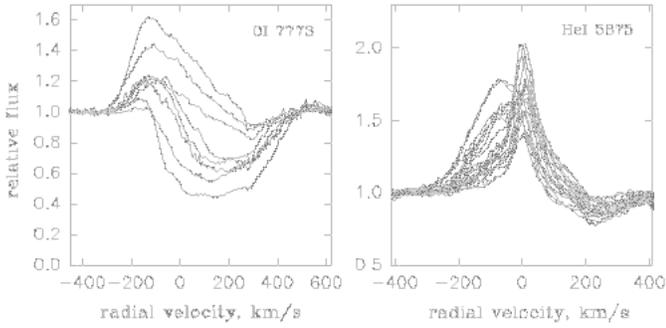}}
\caption{Variability of the inverse P~Cyg--profiles of 
         \ion{O}{i}~7773\,\AA\ and
\ion{He}{i}~5875\,\AA.}
\label{oxi_he1.ps}
\end{figure}

\begin{figure}
\centerline{\resizebox{7.5cm}{!}{\includegraphics{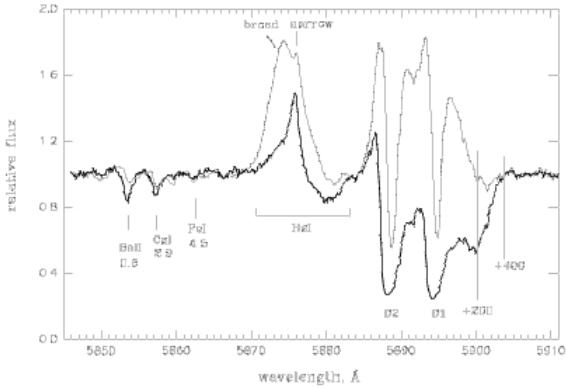}}}
\caption{The two most different spectra show the range of variability in the
region of the \ion{Na}{i}~D and \ion{He}{i} lines. Note the strengthening of 
the \ion{Ba}{ii}
absorption when the accretion components are stronger. The numbers at the
blue lines are the excitation potentials of the lower levels in eV; those
at the \ion{Na}{i} lines indicate velocities relative to D1.}
\label{he1_na1.ps}
\end{figure}

\begin{figure}
\centerline{\resizebox{8cm}{!}{\includegraphics{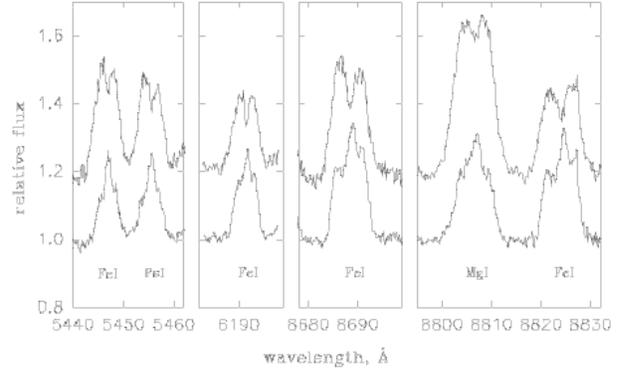}}}
\caption{The narrow absorptions on top of the broad emissions
(upper spectra)
turn into narrow emissions (lower spectra) when the veiling is very high.}
\label{peaks.ps}
\end{figure}

\begin{figure}
\resizebox{\hsize}{!}{\includegraphics{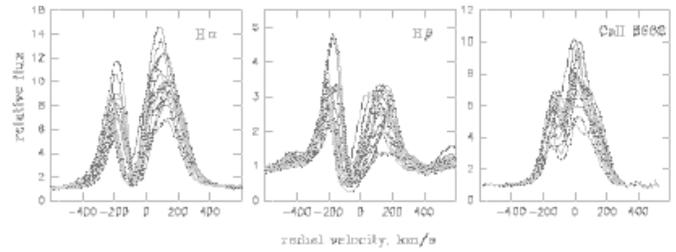}}
\caption{Variability of lines with blue--shifted central absorption
components.}
\label{profil3.ps}
\end{figure}

An outstanding characteristic of the spectrum are the numerous
intensive {\it broad emission lines} (BELs), most of them belonging to
neutral and singly ionised metals. The FWHM of the BELs is
200--280\,km\,s$^{-1}$
for \ion{Fe}{i} and \ion{Fe}{ii} lines, and up to 500\,km\,s$^{-1}$ for H$\alpha$.
The line profiles of the BELs are variable on a time scale of one day.
No large variations of the line profiles were noticed
during 1--2 hours. An example of the BELs' variability is shown in
Fig.~\ref{26all.ps}.
The blue and red wings of the BELs at the line base remain symmetrically
extended to $\pm 200-250$\,km\,s$^{-1}$, while the intensities of the red
and blue parts of the profile can change considerably.

Most of the BELs can easily be identified using the spectrum of the
supergiant 41~Cyg (F5\,II) for comparison (see Fig.~\ref{35avr.ps}).
Note, that narrow absorptions, similar to the WALs, can be found superimposed
on top of the broad \ion{Fe}{i} emissions,
but are usually absent in the \ion{Fe}{ii} emissions.
As a result, the BELs of \ion{Fe}{ii} look split into two parts, while the BELs
of \ion{Fe}{ii} and other ions have a more triangular profile.
These narrow absorptions on top of emission lines vary in radial velocity
and width in correlation with the WALs. One might identify them as the
photospheric lines seen through the optically thin emission. Then, the
absence of these absorptions on top of the \ion{Fe}{ii} lines is 
understandable:
the \ion{Fe}{ii} lines are very weak in the photospheric spectrum of a K dwarf.
However, the
average profile of these absorptions is systematically broader than
that of the WALs.

We measured the equivalent widths of 25 almost unblended, broad emissions of
\ion{Fe}{i} and \ion{Fe}{ii} observed in the spectrum with the most intensive 
lines, and
in the spectrum with the least intensive lines.
Then, the equivalent widths were
converted into fluxes using our photometric data.
In both cases the lines can be brought to a curve of growth
with T$_e$=4300$\pm$500\,K and $\log n_e=9\pm0.5$. However, the lines can be
formed
in non--LTE conditions, and these values should be considered as rough
estimates.

The fluxes in the emission lines show no clear correlation with the
brightness of the star. Instead, there is a good correlation between
the line fluxes and the B--V colour (Fig.~\ref{bv_flux.ps}). This correlation
is partly due to the contribution of the emission lines to the
B and V magnitudes.
For the spectrum with the most intensive emissions, we estimated the
contribution to the B passband as $0\fm4$, and to the V passband
as $0\fm2$. That is, most of the B--V range in Fig.~\ref{bv_flux.ps}
is caused by this effect of the emission lines. The full range of variations
in B--V is, however, much larger (see Fig.~\ref{v_bv.ps}).
Other mechanisms, like temperature
variations or extinction by circumstellar dust may enter.

The next obvious spectral features are the red--shifted absorption components
in many lines. We will refer to these as the {\it accretion components}.
In some lines, like the \hbox{\ion{O}{i}~7773\,\AA\ triplet}, 
\hbox{\ion{Na}{i} D$_1$/D$_2$},
\hbox{\ion{He}{i} D$_3$},
the accretion components
are present permanently, though strongly variable in strength.
The maximum velocity (extension of the red wing) is about 400\,km\,s$^{-1}$.
Examples of these variations are shown in
Fig.~\ref{oxi_he1.ps}.
Note, that the residual intensity at the bottom of the red--shifted
absorption in the oxygen line can be as small as 0.4 of the continuum
intensity.
In Fig.~\ref{he1_na1.ps} we show the two most differing spectra in the region
of the D$_1$, D$_2$ and D$_3$ lines. In the following analysis we will use as
an ``accretion parameter'' the equivalent width (EW) of the D$_1$ absorption
between the velocities of +200 and +400\,km\,s$^{-1}$. There are many other
spectral lines, both neutrals and ions, which occasionally show strong
accretion components.
The spectrum of the accreting gas is described in more detail in
Sect.~\ref{specaccret}. Because of the accretion components,
the maximum intensity
(or centre of gravity) of the BELs is usually blue--shifted.
For example, for \ion{He}{i} broad: --36\,km\,s$^{-1}$; \ion{Fe}{ii}~5316\,\AA:
\hbox{--28\,km\,s$^{-1}$}; Pa\,13: --25\,km\,s$^{-1}$;
\ion{Fe}{i}~5455\,\AA: --14\,km\,s$^{-1}$ and 
\ion{Fe}{i}~6191\,\AA: --8\,km\,s$^{-1}$.

Besides the BELs, there are a few {\it narrow emission lines} (NELs) with
FWHM of about 40\,km\,s$^{-1}$. A comprehensive study of the narrow
emission lines
in spectra of T~Tauri stars (including RW~Aur) was done by
Batalha et al.\ (\cite{Bat96}) with one conclusion being that the lines are
formed near
the magnetic footpoints of the accretion column. In our spectra,
the narrow emission components are clearly visible in \ion{He}{i}~5875\,\AA,
6678\,\AA\ and
7065\,\AA. These lines have
both broad and narrow emission components.
The line profile of \ion{He}{i} can be decomposed into three gaussians:
a broad emission with FWHM=200--250\,km\,s$^{-1}$, centred at
$-40\ldots-50$\,km\,s$^{-1}$, a narrow emission with FWHM=35--60\,km\,s$^{-1}$,
centred at about +10\,km\,s$^{-1}$,
and an accretion component with FWHM=150\,km\,s$^{-1}$, centred at about
+250\,km\,s$^{-1}$.
Only a narrow component is present in the \ion{He}{ii}~4686\,\AA\ emission,
at the average radial velocity of +20\,km\,s$^{-1}$.
Occasionally, weak narrow peaks can be found on top of many other
lines in the spectrum which shows the highest veiling (HJD 2450382.5),
as shown in Fig.~\ref{peaks.ps}. The line D$_3$ is
present in all of our spectra (it falls in the middle of the spectral
order), and therefore we use
it for the analysis of the NEL correlation with other parameters in
Sect.~\ref{periods}.

A blue--shifted absorption component indicating
gas outflow ({\it wind}) is a typical characteristic of
H$\alpha$, H$\beta$, \ion{Na}{i}~D
and the IR triplet of \ion{Ca}{ii}. Examples of variability
in these line profiles are shown in Fig.~\ref{profil3.ps}.

And, finally, {\it forbidden lines} are always present in the spectra,
e.g.\ [\ion{O}{i}]~6300\,\AA, [\ion{S}{ii}]~6716\,\AA \ and 6731\,\AA. 
The line profiles are
similar to those published by Hamann (\cite{Hamann84}).

\section{Veiling versus brightness}
\label{veiling}

\begin{figure}
\centerline{\resizebox{8cm}{!}{\includegraphics{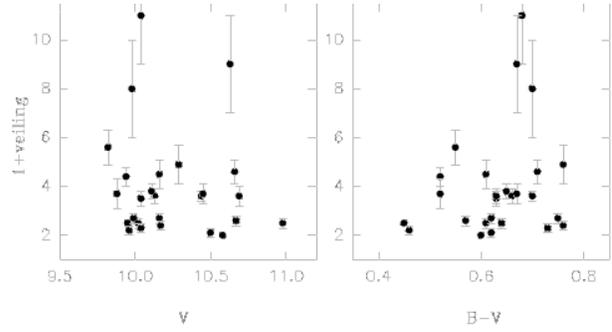}}}
\caption{Veiling versus brightness and colour.}
\label{veil_bv.ps}
\end{figure}

\begin{figure}
\centerline{\resizebox{7cm}{!}{\includegraphics{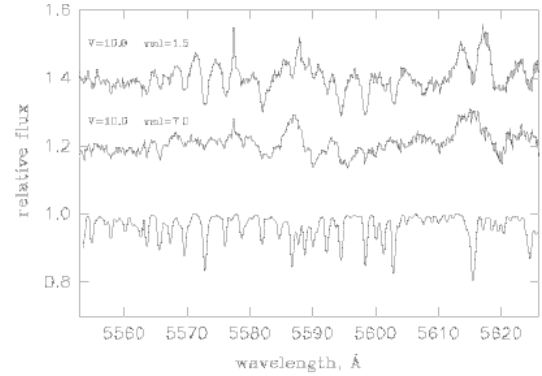}}}
\caption{Two spectra with very different veiling, but with the same
brightness of the star. Lower: the spectrum of $\gamma$~Cep artificially
veiled by a factor 2.}
\label{veil2.ps}
\end{figure}

\begin{figure}
\centerline{\resizebox{6.5cm}{!}{\includegraphics{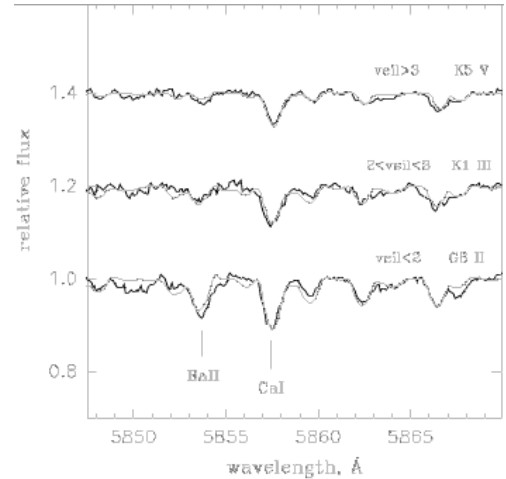}}}
\caption{ Solid lines: average spectra of RW~Aur~A at different levels of
veiling. Thin lines: comparison spectra of 40~Peg (G8\,II), $\gamma$~Cep
(K1\,III--IV)
and 61~Cyg~A (K5\,V), artificially spun up to vsini=30\,km\,s$^{-1}$ and
veiled to the same levels as RW~Aur~A.}
\label{barium.ps}
\end{figure}

The veiling of the photospheric spectrum was determined in
several spectral regions: 5556--5610\,\AA\ (\ion{Fe}{i}, \ion{Ca}{i}),
5719--5751\,\AA\ (\ion{V}{i}, \ion{Fe}{i}), 5855--5870\,\AA\ (\ion{Ca}{i}),
6010--6045\,\AA\ (\ion{Mn}{i}, \ion{Fe}{i}).
In the blue region of the spectrum,
the broad emission lines are blended, and there are no continuum windows.
Therefore we could not select photospheric lines useful for veiling
determinations shortward of 5000\,\AA.

The veiling was derived using two different methods.
The first one is based on the method described by Hartigan et al.\ (1989).
This method consists of a comparison,
within a small wavelength band of a few tens of \AA, between the spectra
of the T~Tauri star with that of
a template star. The latter is chosen so that its spectrum accurately
matches the unveiled
photospheric spectrum of the target. The code implemented in Porto also 
provides
the radial velocity and the rotational broadening of the target star when
compared with the template star. Normally the code adjusts simultaneously
four parameters:
the radial velocity, $v\,\sin i$ and two parameters related to
the continuum excess. The code also provides mechanisms for constraints
on parameter values, useful when some free parameters have been
measured previously using a different method.

For RW~Aur~A we used $\gamma$~Cep (K1\,III--IV) and 61~Cyg~A (K5\,V) as 
template stars. Both templates give similar results within the uncertainties.

The second method is based on measurements of the equivalent widths of
selected absorption lines in spectra of the target and the
template star:
veiling=EW(template)/EW(TTS)--1. The template spectrum was rotationally
broadened to $v\,\sin i=25$\,km\,s$^{-1}$ in order to get the same blends 
of lines as in the spectrum of RW~Aur~A.
In this method we assume that the line width
($v\,\sin i$) is the same in all the spectra, and that the EW of a line varies
due to the veiling variations. The errors come mainly from the
uncertainty in the continuum level, especially in the spectra with
stronger emission lines.

There is a correlation in the time variation of the veiling derived from
different spectral regions. However, the absolute value of the veiling
can differ significantly from region to region. In Table~\ref{Tab1}
we give the average values of the veiling obtained by the two methods
derived from the \ion{Fe}{i} lines in the region of
5556--5610\,\AA. These are the strongest lines of higher excitation
potentials, and they are less sensitive to temperature, which makes
the selection of the template less critical.

The most unexpected result of this study is that {\it no correlation}
was found between the veiling and the brightness of the star,
although the brightness varied by one magnitude, and the veiling
varied over a wide range (Fig.~\ref{veil_bv.ps}). Moreover,
quite different veiling factors
were found in spectra taken at the same brightness of the star, for example,
in the
nights HJD~2451124 and 1125, when photometry was made at two telescopes.
Another example (HJD~2450382 and 0384) is given in Fig.~\ref{veil2.ps}
showing different veilings (1.5 and 7.0) but no significant difference in
brightness. The veiling was determined
in the spectral region of the V band. If the veiling were caused by additional
continuum radiation in this spectral region, the brightness difference
in V would be more than one magnitude!

Another oddity is the variability of ``$v\,\sin i$'' from night to night
in the range 16 to 40\,km\,s$^{-1}$. The effect is real, not instrumental,
e.g.\  the sky line at 5577\,\AA\ remains perfectly narrow in the spectra
which show broadened photospheric lines. The minimal value of $v\,\sin i$
of 16\,km\,s$^{-1}$ can be considered as due to stellar rotation.
No periodicity in the variations of the ``$v\,\sin i$'' parameter was found.
We suggest
that the broader lines are contaminated by an additional absorption from
layers above the photosphere. An argument in favour of this suggestion
can be seen in the variability of the \ion{Ba}{ii}~5853\,\AA\ line. The line was
enhanced in some nights,
that is the ratios of {\ion{Ba}{ii}}/{\ion{Ca}{i}} or 
{\ion{Ba}{ii}}/{\ion{Fe}{i}} were too large for a dwarf.
This strong line, indicating $\log\,g\leq2$ like in supergiants, was observed
mostly in spectra with low veiling (Fig.~\ref{barium.ps}).

We conclude that \ion{Ba}{ii} is formed in a shell above the photosphere.
The same shell must contribute also to absorption in many other lines,
which make them deeper and broader, thus simulating low veiling and
variable $v\,\sin i$. The spectral features of a ``warm shell'' in RW
Aur were also found by Herbig \& Soderblom  (\cite{Herbig80}).
More information about the shell lines is given in Sect.~\ref{specaccret}.

Hence, we conclude that the variations of the veiling we observed in RW~Aur~A
are caused by at least two different processes: 1) the ``true'' veiling due to
continuum + line emission, and 2) absorption in the shell which imitates
low veiling. This may partly explain why the observed ``veiling'' is not
correlated with brightness, although the considerable increase in veiling
at {\it constant} brightness remains a mystery.

\section{Extinction and stellar parameters}
\label{stellpar}

Most of the time RW~Aur~A is probably seen through foreground
circumstellar dust, and it is difficult to derive any precise values of the
interstellar extinction to the star.
We have integrated all FUV spectra obtained
with the International Ultraviolet Explorer (IUE) of RW~Aur~A covering the
spectral region with the interstellar 2200\,\AA\ signature. The integrated
spectrum shows a number of blended emission lines, and we can only set
a rather high upper limit of the average extinction of $A_V \leq 0\fm7$.
Ghez et al.\  (\cite{Ghez97})
give $A_V = 0\fm3$ as a possible upper limit. We have taken one
spectrum of RW~Aur~B, $1\farcs4$ from A, showing only weak traces of
interstellar Na~D absorption. The total equivalent width of these
lines amount to 0.2\,\AA, also consistent with a low interstellar
extinction to RW~Aur. Hence, we assume that the
interstellar extinction to RW~Aur is low, which is consistent with its
location outside any molecular cloud boundary.

When the star is brightest and bluest, we expect the circumstellar
extinction to be minimal.
A minimum of  \hbox{B--V=$0\fm45$} was observed on HJD~2450386 (see
Table~\ref{Tab1}).
The colour is too blue for a K dwarf, indicating the presence of a
hotter continuum source also responsible for the veiling.
On this occasion, the star had V=$9\fm95$ and veiling=1.5. Taking into account
the correction for the contribution of emission lines ($0\fm2$ in V) and
the correction for the veiling ($2.5\,\log(1+{\rm veiling})$), we find a
corrected V=$11\fm14$. With the distance of 140\,pc (Elias
\cite{Elias78}) and $A_V= 0\fm3$, the
absolute magnitude is $M_V= 5\fm11$.
Using a bolometric correction appropriate for the spectral types
\hbox{K1--K4},
which is BC$= 0\fm2 \ldots 0\fm5$, we get $M_{\rm bol}=4\fm9...4\fm6$,
which corresponds to $L/L_{\sun}=0.85 \ldots 1.1$.
The derived luminosity represents a lower limit, since circumstellar extinction
can be present.

From the spectral type range and the luminosity we can estimate the radius
of the star as $1.3 \ldots 1.5\,R_{\sun}$.
Since the luminosity is a lower limit, the radius is also a lower limit.

With the lower limit of the luminosity, and with the
range in spectral type, we obtain a lower limit for the mass of the star of
about 1.1\,$M_{\sun}$
according to evolutionary model tracks (e.g.\  Palla \& Stahler
\cite{Palla99}).
Notice, that with these values of mass and radius, the free--fall velocity
(starting from a distance of 10 stellar radii) is about 440\,km\,s$^{-1}$ at
the stellar surface. This is in a good agreement with the observed maximum
accretion velocity of 400\,km\,s$^{-1}$.

\section{Periodicities in spectral line variations}
\label{periods}

\begin{figure}
\centerline{\resizebox{7.5cm}{!}{\includegraphics{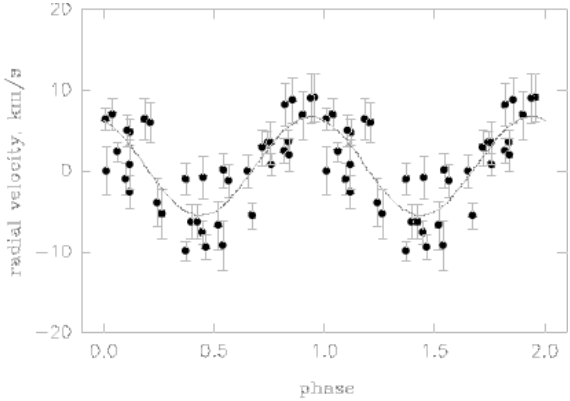}}}
\caption{Phase diagram of the radial velocity variations
of the weak absorption lines, \hbox{$P=2\fd7721$}.}
\label{wal_vr.ps}
\end{figure}

\begin{figure}
\centerline{\resizebox{7.5cm}{!}{\includegraphics{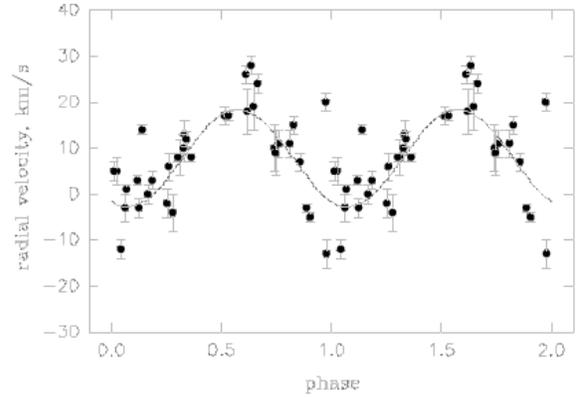}}}
\caption{Phase diagram of the radial velocity variations
of the \ion{He}{i}~5876\,\AA\ narrow emission, \hbox{$P=2\fd7705$}.}
\label{he1_vr.ps}
\end{figure}

\begin{figure}
\centerline{\resizebox{6cm}{!}{\includegraphics{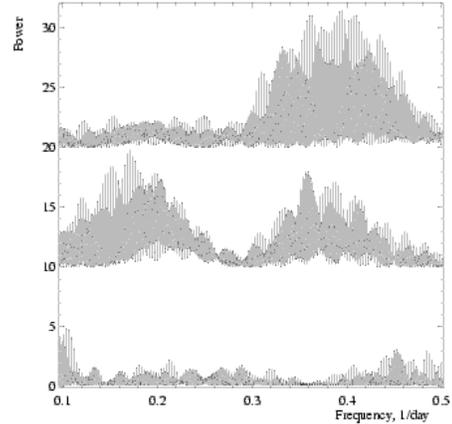}}}
\caption{Power spectra: upper --- 
 EW of the narrow emission of \ion{He}{i}~5876\,\AA,\newline
centre --- EW of the broad emission of \ion{He}{i}~5876\,\AA,\newline
lower --- spectral window.}
\label{power.ps}
\end{figure}

\begin{figure}
\centerline{\resizebox{6cm}{!}{\includegraphics{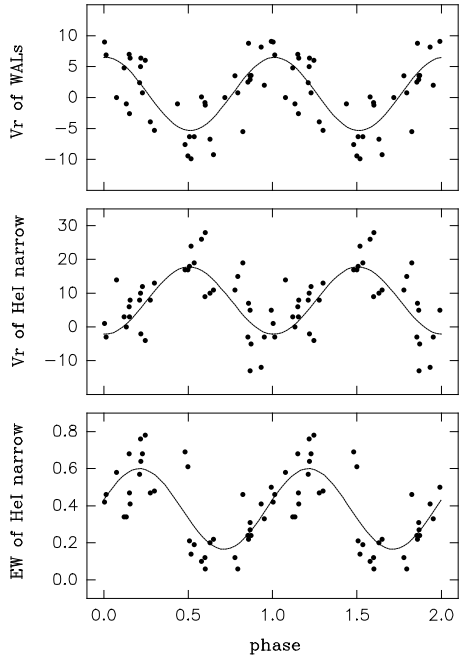}}}
\caption{Phase shifts between the radial velocities of the WALs,
the radial velocities of the \ion{He}{i} NELs
and EWs of \ion{He}{i} NELs, \hbox{$P=2\fd7710$}.}
\label{wal_he1.ps}
\end{figure}

\begin{figure}
\centerline{\resizebox{6cm}{!}{\includegraphics{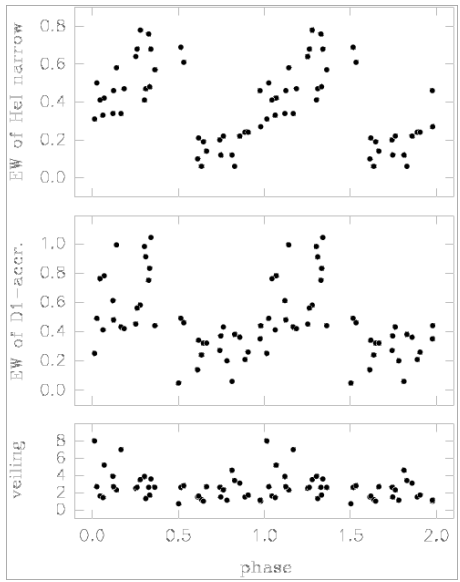}}}
\caption{Phase diagrams of the EW of the \ion{He}{i} narrow emission 
(upper panel),
the accretion components of \ion{Na}{i}~D$_1$ (middle panel)
and the veiling (lower panel), \hbox{$P=2\fd7705$}.}
\label{he1_d1.ps}
\end{figure}

\begin{figure}
\centerline{\resizebox{4cm}{!}{\includegraphics{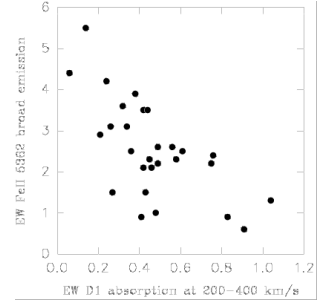}}}
\caption{Anti--correlation of the accretion components of \ion{Na}{i}~D$_1$ and
the EW of the broad emission of \ion{Fe}{ii}~5362\,\AA.}
\label{d1_fe2.ps}
\end{figure}

In Paper~I, we have reported already the discovery of periodic
variations in the photospheric and emission lines. Since then
we have collected more spectra in 1999; the new spectra confirm
that the period and the phase remain stable.
All radial velocities were measured by cross--correlation with the
template spectrum of $\gamma$~Cep (K1\,III--IV) in the regions
5550--5610\,\AA\ and 6000--6050\,\AA.
The average radial velocity
derived from the photospheric spectrum is +16\,km\,s$^{-1}$
and is subtracted from all measured heliocentric velocities.
The radial velocities of the WALs vary from --6 to +6\,km\,s$^{-1}$. A
periodogram
analysis indicates a period most likely in the range from $2\fd5$ to $2\fd9$.
There is a number of almost equal, equidistant peaks,
separated in frequency by 1/year.
The strongest peak is at $P=2\fd7721$. A phase diagram  with this
period is shown in Fig.~\ref{wal_vr.ps}.

In addition to the WALs, we also analyzed the periodicities in other groups
of spectral features:  NELs, BELs and accretion components.
We found that the WALs, NELs and accretion components all show about
the same set of frequencies in the power spectrum, with a period in
the range of $2\fd5$ to $2\fd9$. The BELs show a twice longer period in the
range of 5 to 6 days. The period around $2\fd77$
is also present in the variations of the BELs, but with less power than
the 5--6 days period.
This is illustrated in Fig.~\ref{power.ps}.

With our set of data we cannot prove that there is only one and the same
period in the variations of the WALs, NELs and accretion components.
For example, the power spectrum of the NELs shows that the most
significant peaks are slightly shifted with respect to those of the
WALs. The best period for the
radial velocities of the narrow \ion{He}{i} line is $P=2\fd7705$ (compared to
$2\fd7721$ for the WALs). The corresponding phase diagram is shown in
Fig.~\ref{he1_vr.ps}.
Note the systematic shift in the mean radial velocity of the \ion{He}{i} NEL
by about +8\,km\,s$^{-1}$.
The velocity amplitude is larger than that of the WALs.

The EW of the \ion{He}{i} NEL varies from 0.15 to 0.60\,\AA\ and
shows the same periods as the radial velocity, with the strongest
peak at $P=2\fd7705$. In order to check if the WALs and the \ion{He}{i}
emission vary in phase or with a phase shift, one must plot a phase diagram
with a common period. A compromise period of $P=2\fd7710$
was used for the diagram shown in Fig.~\ref{wal_he1.ps}.
The result is that the radial velocity of the WALs and the \ion{He}{i} NEL
vary in anti--phase (phase shift 0.5),
while the EW of the \ion{He}{i} NEL shows a phase shift of about 0.25.
Any other common period in the $2\fd5$ to $2\fd9$ interval shows
the same phase shifts.

The \ion{He}{ii}~4686\,\AA\ has only a narrow emission component, and varies
in correlation with the narrow emission in \ion{He}{i}~5876\,\AA. However,
the mean radial velocity is even more red--shifted, +20\,km\,s$^{-1}$.

The EWs of the accretion components vary in phase with the
EW of the \ion{He}{i} narrow emission. This is shown in Fig.~\ref{he1_d1.ps}
for the \ion{Na}{i}~D$_1$ red--shifted absorption between +200 and 
+400\,km\,s$^{-1}$.
The stronger the \ion{He}{i} NELs, the larger is the accretion absorption.

The EWs of the broad emissions of \ion{He}{i}, \ion{Fe}{i}, \ion{Fe}{ii}
vary in correlation with each other. An anti--correlation exists
between the variations of the EWs of the accretion components
and the EWs of the broad emission lines of \ion{He}{i}, \ion{Fe}{i}, 
\ion{Fe}{ii}
(Fig.~\ref{d1_fe2.ps}). More about correlations between different
spectral features is given in the next section, where the spectral
line profiles formed in the accreting gas are discussed.

No periodicity was found in the veiling variations, and there is no obvious
correlation between the veiling and any other spectral feature.
Only a weak correlation exists between the veiling and the
EW of the \ion{He}{i} narrow emission.
Conversion of EWs into fluxes does not improve the correlation.

\section{The spectrum of the accreting gas}
\label{specaccret}

\begin{figure*}
\centerline{\resizebox{12cm}{!}{\includegraphics{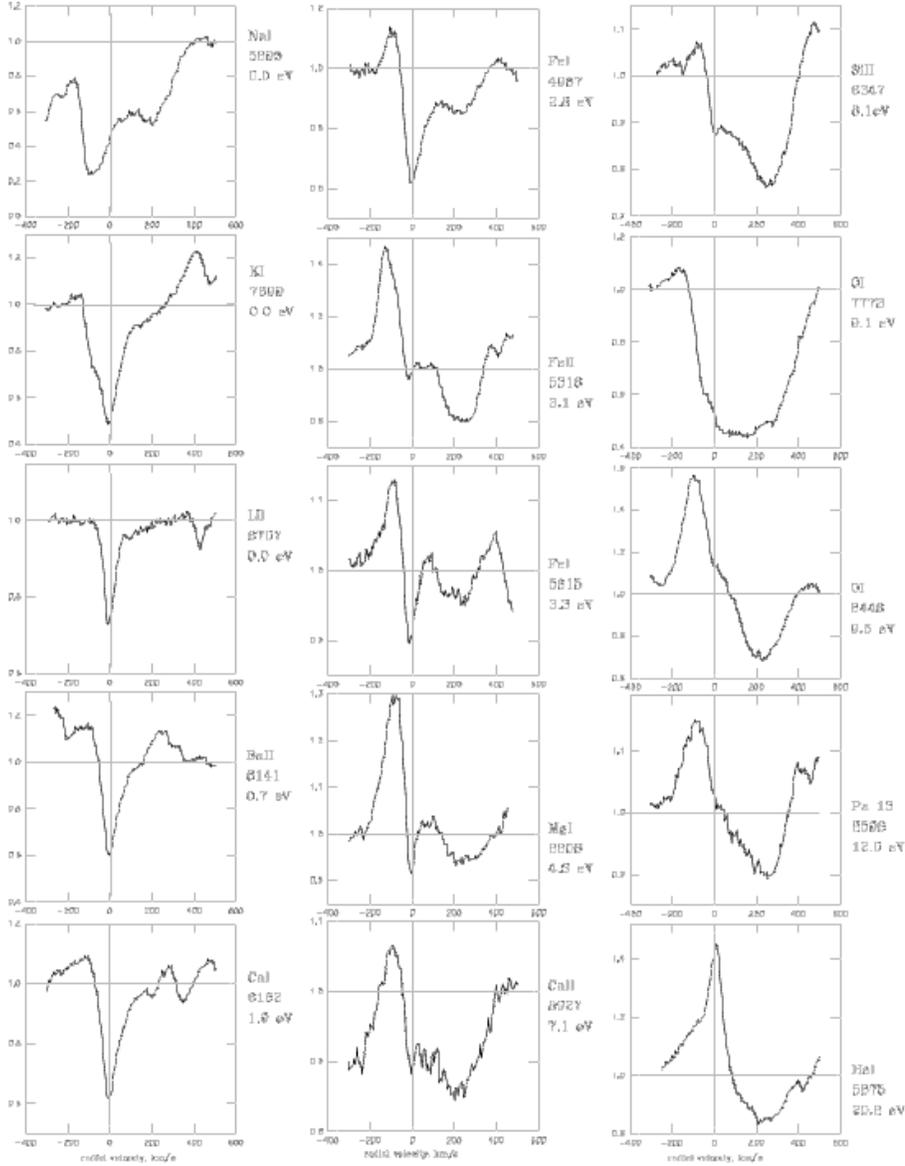}}}
\caption{Accretion components of lines of different excitation potentials
in the spectrum with enhanced accretion (HJD 2450388.5).}
\label{accr.ps}
\end{figure*}

The best indicators of accretion are the lines of \ion{Na}{i}~D and the
\ion{O}{i}~7773\,\AA\ triplet.
Occasionally, when accretion is enhanced, red--shifted components
appear in many other lines. In one night (HJD 2450388.5)
the accretion was exceptionally strong. The accretion components
were present in almost all the emission lines, which gives an opportunity
to describe the spectrum of the accreting gas in more detail.
Profiles of selected lines of different excitation potentials are shown
in Fig.~\ref{accr.ps}. The common feature is that the accretion line profile
consists of two different components: a low velocity absorption at about
0\,km\,s$^{-1}$ or slightly blue--shifted, and a high velocity absorption
with the bottom  around +250\,km\,s$^{-1}$ and extending to +400\,km\,s$^{-1}$.
The low excitation lines of \ion{K}{i}, \ion{Li}{i}, \ion{Ba}{ii}, 
\ion{Ca}{i}  and \ion{Fe}{i} all show
enhanced low--velocity absorptions with a weak red--extending wing.
The lines of excitation potentials between 2 and 7\,eV have both low--velocity
and high--velocity absorptions. The lines of the highest excitations,
more than 8\,eV, show mostly high--velocity components.

The presence of accretion--induced enhancements of the low excitation,
low ionization
lines in the spectrum of RW~Aur~A were reported recently by Stout--Batalha et
al.\ (\cite{Stout00}) with emphasis on \ion{Li}{i}. They suggested that the
enhancement of absorption at the line centre is due to additional
absorption by cool gas at the
beginning of the accretion flow.

\begin{figure}
\centerline{\resizebox{4cm}{!}{\includegraphics{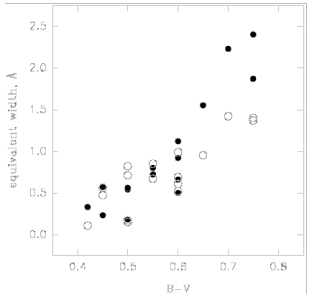}}}
\caption{Correlation of the EW of the low velocity absorption in
\ion{K}{i}~7698\,\AA\ (filled dots) and of the EW of the central absorption in H$\beta$
with B--V.}
\label{k1_beta.ps}
\end{figure}

The EW of the central absorption of \ion{K}{i}~7699\,\AA\ can be used as an 
indicator
of this low velocity gas. We find a correlation between the EW of
\ion{K}{i}~7699\,\AA\ and the  EW of the {\it blue}--shifted absorption of 
H$\beta$,
both correlating with B--V (Fig.~\ref{k1_beta.ps}).
This is an observational evidence that {\it the inflow and the outflow are
related processes}.

\begin{figure}
\centerline{\resizebox{7.5cm}{!}{\includegraphics{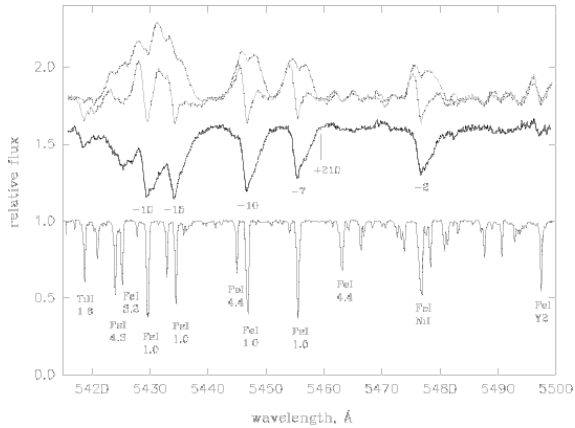}}}
\caption{ The uppermost two spectra of RW~Aur~A are the average spectrum
          and the spectrum with enhanced accretion (dashed). Below them, their
        difference (the differential spectrum) is shown with radial velocities
        indicated. At the bottom we
        show the spectrum of $\gamma$~Cep (K1\,III--IV) with line 
        identifications and excitation potentials.}
\label{shell42.ps}
\end{figure}

The difference between this spectrum with enhanced accretion and
the average spectrum of RW~Aur~A is shown in Fig.~\ref{shell42.ps}.
Note that the dips of the absorptions are blue--shifted from $-2$ to
$-15$\,km\,s$^{-1}$. The shift is larger for stronger lines (up to $-100$
km\,s$^{-1}$ in Na~D). No molecular bands of TiO were found in the
differential spectrum.

In several lines of neutrals and ions we measured the
relative depths of the low velocity components and the relative depths at
+250\,km\,s$^{-1}$. Assuming LTE conditions we find T$_e= 4700\pm 300$\,K,
$\log n_e=9.2\pm 0.6$ for the low velocity components, which are similar to
the parameters obtained for the broad emission lines.
The high--velocity absorptions are
formed at higher temperature ($>8000$\,K) and higher densities
($\log\,n_e > 13$) but different sets of lines give different values.

\section{Discussion}
\label{discussion}

The extremely complex, yet regular patterns of spectral variability together
with the ``wild'' photometric behaviour of RW~Aur~A, challenge any attempt
to model the star and its surrounding. 
The main result is the periodic,
anti--phase variations in the radial velocities of the WALs and NELs. Such
a periodicity in principle can be induced either by {\it orbital} motion of an
invisible, low--mass secondary (see Paper I), which also influences the
gasflows around the star, or by {\it rotational} modulation of a single star
with an inclined or asymmetric magnetosphere.
In the following, we present
two sketches of models, which we hope can form a basis
for further developments, and also provide indications of future observational
tests.

Both models are critically dependent on the inclination of the stellar
rotation axis and a possible stellar magnetosphere.
In Paper~I we discussed how the radial velocity measurements of the jets at
RW~Aur~A can be used to restrict the range of possible orientations.
Recently, C.~Dougados (private communication) has communicated
to us a preliminary estimate of the inclination of the
counter--jet of the star, based on the proper motion and radial velocity of
one knot in the jet. The inferred inclination (to the line--of--sight)
is 67$\pm$4 degrees (see also Dougados et al.\ \cite{Dougados00b}).
We adopt this value as the direction of the jet. (The gas could move
faster than the knots, in which case the value is a lower limit).
In addition we will assume that the jet axis is perpendicular to
a circumstellar dusty disk surrounding RW~Aur~A, and parallel to the
axis of stellar rotation.

We need to explain the following observed properties:
\begin{enumerate}
\item Periodic, sinusoidal variations of the radial velocities of the WALs
and NELs, and, in addition, of the
equivalent widths of the NELs and the accretion components, all with
a period of $P=2\fd77$.
\item Anti--phase  between the WAL and NEL radial velocities.
\item The total amplitudes of the \ion{He}{i} and \ion{He}{ii} NEL 
radial velocities are
equal, but the average velocities are redshifted with respect to the
average WAL velocity --- \ion{He}{ii} significantly more than \ion{He}{i}.
\item A $\sim$1/4 phase--shift between the variations of the equivalent
widths and radial velocities of the \ion{He}{i} NELs.
\item Correlation between the NEL and the accretion component equivalent
widths, with a burst of accretion occurring during each cycle.
\item Periodic variations of the equivalent widths of the BELs with about
the double period, $\approx 5\fd5$.
\item Anti--correlation between the BEL and accretion component equivalent
widths.
\end{enumerate}

Common for both models is that the observed variabilities of the
red--shifted absorptions are related to the stream(s) of gas channeled by
a magnetic field to hot spot(s) on the stellar surface.
Stellar activity alone cannot account for all the details:
{\it non - axisymmetric accretion} is present.

The broad emission lines with the blue--shifted maximum and red--shifted
absorptions are most probably formed in a global magnetosphere threaded by
streams of gas flowing towards the star (Calvet \cite{Calvet98} and references
therein, Muzerolle et al.\ \cite{Muz98}). The free--fall time is much less 
than the period of rotation, i.e.\ local fluctuations in the accretion
flows would result in irregular variations of the blue and red wings in
the broad emission line profiles.

In projection onto the star, along the same line of sight, we see the low
velocity, low temperature gas at the beginning of the accretion column,
and the high velocity, high temperature gas approaching the star
at the end of the accretion column
(Fig.~\ref{accr.ps}).
The narrow emission lines of He can be formed within the same accretion
column but very near the stellar surface, where the gas is already
decelerated but still has some positive velocity. This explains the
shifts in the mean radial velocity of NELs, and the correlation between
the EWs of the NELs and the accretion components
(Fig.~\ref{he1_d1.ps}).
These
narrow \ion{He}{i} emission lines can be used for the detection of the magnetic
field of this star, as was made for BP Tau by 
Johns--Krull et al.\ (\cite{JohnsKrull99}).

The lines of \ion{Ba}{ii} discussed in
Sect.~\ref{veiling}
and the enhancements of the
low excitation lines of \ion{Li}{i}, \ion{K}{i}, \ion{Ca}{i} originate 
probably in the outer part of the magnetosphere, which acts like a shell.
The strengths of these absorptions correlate with B--V,
which may be due to the presence of dust in dense blobs of gas rising from
the accretion disk. The blue--shifted "dips" in these lines
indicate that a wind component is also present in the profile.
Presumably, the ultraviolet blue--shifted absorption components
reported by Gahm (\cite{Gahm70}) and Errico et al.\ (\cite{Errico00}) 
originate in the same wind outside the main accretion flows.

Finally, the light variability most certainly is due to variations
both in circumstellar extinction and the continuous veiling.

The properties described above are common for both models, but the
origin of the periodical modulation is different.

\subsection{ Binary star?}

\begin{figure}
\centerline{\resizebox{5.5cm}{!}{\includegraphics{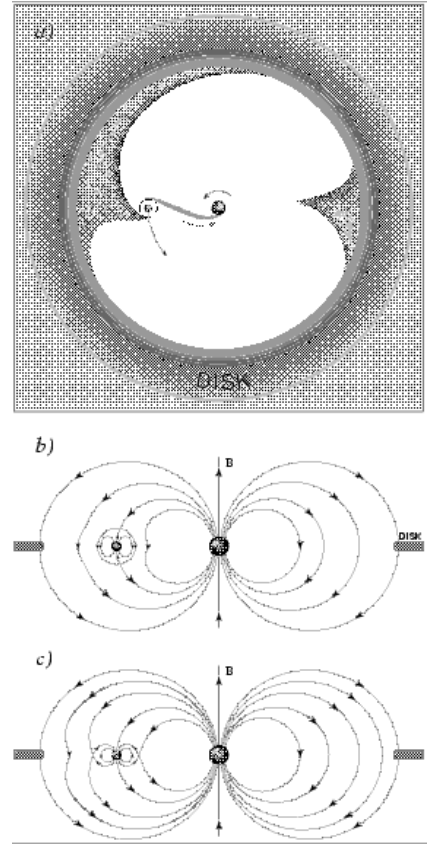}}}
\caption{In our search for mechanisms explaining {\it how} the secondary may
  act as a generator of enhanced accretion from an area at the disk
  edge to a spot on the primary we discuss gravitational perturbations
  leading to accretion along spirals to the secondary, where some
  matter is deflected and continues to fall down to a spot, possibly along a
  trajectory tilted relative to the primary surface (upper panel).
  Another concept is described in the lower two panels, showing the
  combined magnetic field structure of two magnetic stars. The field
  at the secondary looks very different if the magnetic bodies have
  parallel (middle panel) or anti--parallel fields (lower panel). Enhanced
  MHD action occurs at the disk edge closest to the
  secondary.}
\label{ggsketch.ps}
\end{figure}

\begin{figure}
\centerline{\resizebox{6.5cm}{!}{\includegraphics{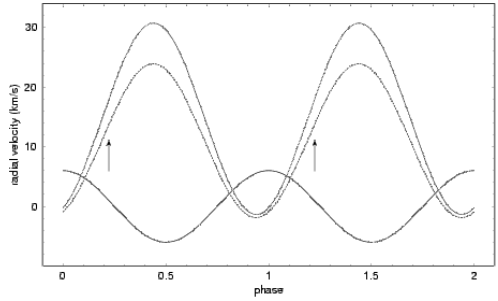}}}
\caption{Calculated radial velocity changes
         of the WALs (fulldrawn), the He\,I NELs (dashed),
         and the He\,II NEL (short--dashed). The phase whith expected
         maximum in the EWs of He are marked with arrows.}
\label{vnel.eps}
\end{figure}

Paper~I addressed several aspects of the binary hypothesis, where it
is assumed that the WAL velocity variations are due to orbital motion. First
of all it is reassuring that our new observations confirm the period, amplitude
and phase of the radial velocity variations of the WALs. In addition, we
now have the inclination of the system
(assuming that the orbital axis is aligned with the jet).
We expect that the K--type primary has a mass slightly
larger than 1\,$M_{\sun}$
(see Sect.~\ref{stellpar}).
It follows that the secondary is a brown dwarf (BD) of
$0.04-0.05\,M_{\sun}$ in an orbit of low
eccentricity at about 8--10\,$R_{\sun}$ from the primary (see Paper~I).
The orbital period of the secondary is $2\fd77$, and the centre of
gravity is well inside the primary.

The basic concept of the binary star model is that the periodic
spectral variability is related to the influence of a close secondary
component, just like for other close binary TTSs. The \ion{He}{i} and
\ion{He}{ii}
NELs cannot originate at hot spots on the
{\it secondary} unless the inclination of the system is very small (Paper~I).
Moreover, this hot region cannot be located {\it between} the components,
for instance as the result of two colliding winds,
because then the amplitudes of the NEL
radial velocities would be much larger than observed. However, many
TTSs show narrow \ion{He}{i}
emission components thought to arise
at the footpoints of {\it global}
accreting flows from the disk, and we have followed this general concept.

Consequently, our discussion focusses on the following idea: global
magnetospheric accretion, typical for TTSs, is present. The BELs are
formed in this flow. In addition, the secondary produces enhanced
accretion in one direction. This gas falls freely into regions close to the
stellar surface of the {\it primary}, where deceleration and heating
occurs. We have considered two possible mechanisms, which may lead
to such an asymmetry of the accretion. Both these possibilities are shown 
schematically in
Fig.~\ref{ggsketch.ps}, where the disk edge is placed at the 2:1 period
resonance (i.e., the orbital period of the inner part of the disk is
about $5\fd5$).

One possibility is that the secondary, which has cleaned a gap in the
disk, also perturbs the disk. Gas spirals from the disk,
preferentially to the secondary, where some matter is deflected and
falls onto the primary, possibly along spiral trajectories.  Similar
cases have been calculated by e.g.\ Artymowicz \& Lubow
(\cite{Art96}), and could explain the periodic bursts of emission
observed in TTSs with close massive components in elliptic orbits,
like DQ~Tau (Basri et al.\ \cite{Basri97}). However, we envision a
secondary in a nearly circular orbit, and it is not massive. This
particular case must be followed numerically, assuming an upper limit
of the ellipticity of 0.2 (according to Paper~I), before any
conclusions about plausibility can be made.

Another possibility is that both stars are
magnetic, and that the magnetospheres interact in such a way that
enhanced accretion from the disk occurs along dipole
magnetic field lines to the primary in the direction to the secondary.
In this case, the secondary receives rather little of the accretion
flow, and acts only as a generator of enhanced MHD action in a small
region of the disk edge.  Carlqvist (private communication) has provided
the principles for a combined magnetic field from two magnetic stars (dipoles)
based on early magnetospheric models. For the
configurations in Fig.~\ref{ggsketch.ps}, the surface magnetic field strength
of the contracting BD (radius $\sim 0.3\,R_{\sun}$) was assumed to be 
about 1/10 of that of
the primary. These two cases represent parallel and anti--parallel field
orientations. Note, that for enhanced MHD action to be efficient, the
disk radius must be smaller or the BD magnetic field stronger than
assumed in Carlqvist's calculation.
The recent discovery of a large
flare on a brown dwarf by Rutledge et al.\ (\cite{Rutledge00}) shows that
the magnetic field can be strong.

We have considered several possible models for how gas is channeled
into a cylinder over an area (``spot'') on the primary. The narrow \ion{He}{i}
and \ion{He}{ii} lines are formed as a result of heating from an impact shock.
When the secondary
is closest to us, also the spot is facing us, since maximum lineflux
occurs at a phase around 0.25.
The spot must also be seen all the time, and
is therefore located at high latitude. A similar spot may exist on the
``southern'' stellar hemisphere, but is tilted out--of--view. In addition,
the He lines have maximum positive velocities at phase 0.5, implying
that the final impact must occur along trajectories trailing
the secondary.

Fig~\ref{vnel.eps} shows one example of calculations using a simple
geometrical model. In this model, the centre of the primary,
the spot, and the secondary
are orbiting the centre of gravity. The spot co--rotates at high latitude,
where the orbital motion of the spot is
small. The infall is nearly
radial to the normal of the surface, but strongly
tilted in longitude. In this way the anti--phase
variations of the WALs and NELs can be reproduced, as well as
the amplitudes of the NELs. Assuming that the EWs of He are largest at
maximum projected spot area, the  corresponding observed $\sim$1/4 phase--shift
can also be
reproduced. The NELs of \ion{He}{i} and \ion{He}{ii} are assumed to form 
along the same trajectory but with different
average infall velocities (tracing different volumes in the cylinder).
The shift in average velocity in Fig~\ref{vnel.eps}
corresponds to infall velocities of 30\,km\,s$^{-1}$ (\ion{He}{i}) and 
40\,km\,s$^{-1}$ (\ion{He}{ii}).

In the binary case, part of the veiling comes from a ring around
the visible pole connected to global accretion as manifested in the BELs,
just as
assumed for other TTSs. Superimposed is the veiling produced in the ``spot''.
It could be that this spot is a region of enhanced
accretion in the high latitude ring. Both these components can be
expected to vary irregularly and independently due to variable mass
inflow. Part of the veiling may also be due to narrow emission lines,
as is discussed in the subsequent section.

Referring to the model
descriptions above it seems worth exploring accretion
paths in the combined magnetospheres of the stars. Since the disk edge
is located far outside the co--rotation radius, the global field is
dragged behind, in a way similar to the solar magnetic field. The field lines
can therefore be expected to be curved also in longitude.
The magnetosphere of the secondary acts as the motor of
channelled accretion in this case.

In conclusion, it seems that the binary hypothesis can account for
most of the points set above. Many of the details
are presently not very well understood, and require a much deeper analysis of
the physics involved. A major item is point 6, namely the $5\fd5$
period, which calls for an additional modulation of the BELs.

\subsection{ Single star?}

Periodic variations in the accretion features are not unique
to RW~Aur~A, and were observed also in other TTSs, like SU~Aur 
(Giampapa et al.\ \cite{Giam93}, Petrov et al.\ \cite{Petrov96}).
As an explanation it
was proposed that the periodicity is related to the {\it axial rotation}
of the star. If the magnetic axis is not aligned with the rotational
axis, the accreting gas gives rise not to hot rings, but to two
elongated hot spots near the two magnetic poles of the star
(K\"onigl \cite{Konigl91}).
Then, the stellar rotation would modulate the brightness and the spectral
features associated with the accretion. The same effect of the rotational
modulation may appear if the structure of the stellar magnetic field is not
perfectly symmetric, with larger magnetic loops at certain
longitudes. In both cases, we have non--axisymmetric magnetospheric
accretion. Note that permanent active longitudes have been found in other
types of active stars (Jetsu et al.\ \cite{Jetsu93}; Berdyugina \&
Tuominen \ \cite{Berd98}).

The basic concept of the single star model of RW~Aur is that the longer period
of 5--6 days is the rotational period of the star, while the shorter one,
about $2\fd77$, indicates that there are two active regions on the
stellar surface --- two footpoints of the accretion columns on opposite
sides of the star.

The following arguments support this concept. With $R_*\geq 1.3 R_{\sun}$
(Sect.~\ref{stellpar}) and $v\,\sin i=16$\,km\,sec$^{-1}$
the period must be $\geq 4\fd1$. That is, the observed 5--6 days period
is more likely to be the true rotational period than the $2\fd77$ period.
A similar period of $5\fd4$ was first found in variations of
the H$\beta$ profile
in RW~Aur~A in observations of 1980--82 (Grinin et al.\ \cite{Grinin83}).
Variations in the broad emission lines formed in a large volume of
the magnetosphere and the stellar wind reflect the rotational period,
while the spectral features associated with the accretion
processes closer to the star reflect the local structure of
the accretion channels and may show period(s) shorter than the rotational
one.

\begin{figure}
\resizebox{\hsize}{!}{\includegraphics{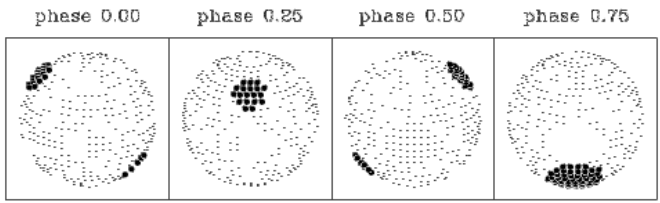}}
\caption{A model of a star with two spots of chromospheric emission (solid
dots), associated with the two footpoints of the accretion columns. The
photospheric absorption spectrum contributes in each dot over the
entire surface, while narrow emission lines only arise in the spot
area.}
\label{phases.ps}
\end{figure}

\begin{figure}
\centerline{\resizebox{7.5cm}{!}{\includegraphics{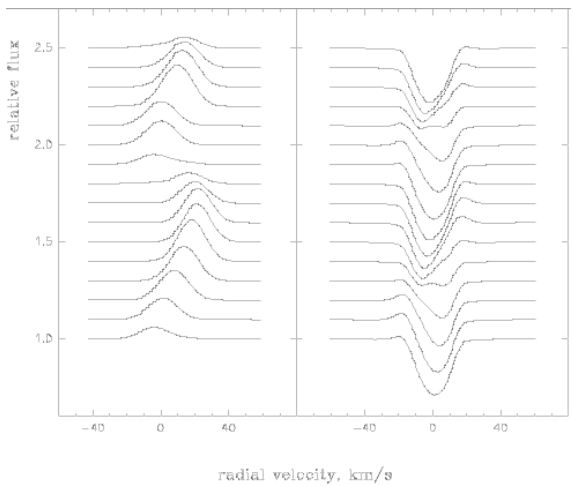}}}
\caption{Variations of emission and absorption line profiles in the spectrum
of the spotted star (shown on Fig.~\ref{phases.ps}) , integrated over the
stellar surface. Each spectrum corresponds to a certain phase of the
axial rotation; step in phase = 1/16.}
\label{lines16.ps}
\end{figure}

The phase--shift in the variations of the radial velocities  and the EW
of the \ion{He}{i} narrow emission, about 0.25, is a clear indication that the
emission is formed in a spot--like area. The EW is larger when the spot is
face--on to the observer, while the radial velocity is larger when the spot
is near the limb. With two spots on opposite sides, this happens twice over
one rotational period of about $5\fd5$, hence the period of $2\fd77$
appears in the data.

The unexpected result of our research is the absence of any
correlation between veiling and brightness, although both
parameters were changing over wide ranges.
The considerable increase of veiling at constant brightness of the
star (Fig.~\ref{veil2.ps})
excludes the possibility that the veiling was caused by a rise in
the additional continuum radiated by a hot spot (unless the increased
brightness was precisely balanced by increased circumstellar extinction).

Besides the hot continuum, the veiling can be raised by narrow
emission lines of neutral metals which partially fill in the photospheric
absorptions. Variability in the strength of these hypothetical narrow
emissions could cause variability of the veiling even at constant
brightness of the star.
An argument in favour of such an interpretation can be found
in the spectrum taken at the highest value of the veiling (HJD~2450382.5).
As was written in
Sect.~\ref{Multicomp},
narrow absorption components are
usually present on top of many broad emission lines of neutral metals.
In the spectrum taken at highest veiling, these narrow absorptions
turned into narrow emissions (Fig.~\ref{peaks.ps}).
The radial velocities of these narrow emissions  of the metals
correspond to that of the \ion{He}{i} narrow emission. We may suggest that the
narrow emissions of the metals form in a chromospheric--like region
associated with the
footpoint of the accretion column. Since we know already that the radial
velocity of the \ion{He}{i} narrow emission varies periodically, we may expect
that the narrow emissions of the metals also vary in radial velocity,
and this must be reflected in the variability
of the photospheric line profiles distorted by these narrow emissions.

In order to check this hypothesis numerically, we calculated a synthetic
spectrum from a single rotating star with two spots of ``chromospheric''
emission. Fig.~\ref{phases.ps} shows the positions of the two spots as
they are projected onto the stellar disk at different phases during
the rotation of the star. Fig.~\ref{lines16.ps} shows an example of
line profile variations in an absorption line partially filled in
with the chromospheric emission, and in a pure emission line (like HeI).
As the star rotates, the radial velocities of the absorption line vary
in anti-phase with respect to those of the emission line, and the EWs of
the emission vary with a phase-shift to the radial velocity, similar
to that observed in RW~Aur. The observed amplitudes and phase shifts can
be reproduced with two spots of about 20 degrees in radius, located at
latitudes of $\pm45^{\circ}$ at opposite longitudes, as is shown in
Fig.~\ref{phases.ps}. With an inclined magnetic axis the spots may be close
to the magnetic poles.
One may notice, however, that the velocity variations in such a model are
not quite sinusoidal.
Spectra with higher signal--to--noise ratios are needed to check whether the
predicted absorption line distortions shown in Fig.~\ref{lines16.ps} are
really present.
Although the hypothesis of the chromospheric emissions filling in the
photospheric lines accounts for most of the observed variations, it is
not clear why these emissions were never seen above the continuum level
in all the photospheric lines.

The hot spots in CTTSs are not as long--living as the cool spots
related to the {\it local} magnetic fields, i.e.\ accretion processes are
not stable. Contrary to the case of a binary star, the variations
in the radial velocities caused by the spot effect described above
may disappear sometimes. This could be an observational test to distinguish
between the two models.

\section{Conclusions}

Our spectroscopic and photometric observations of the
unusual T~Tauri star RW~Aur~A have revealed extremely complex patterns
of variability in all emission and absorption lines, both with regard to
line profiles, equivalent widths/fluxes and radial velocities of the
spectral components. A number of these components vary with a period
of $2\fd77$. The variations are regular, and indicate that these patterns
have been stable over several years.

In Paper~I we reported the discovery of periodic, sinusoidal, small
amplitude radial velocity changes in the weak spectrum of narrow
absorption lines (WALs). Our new observations in Nov.\ 1999 confirm
the period, phase and amplitude of the WAL velocities.
Relative to these, the narrow emission lines of He vary in anti--phase.
The broad emission lines, dominating in the spectrum of the star,
show also the double period of 5--6 days. The veiling shows no corelation
with brightness, although both parameters vary in a wide range.
This is partly due to additional variable opacity of the layers
above the photosphere (the shell), which imitates lower veiling
and also makes the spectrum of RW~Aur~A vary in spectroscopic
luminosity. The presence of chromospheric-like line emission in the
photospheric lines most likely also plays a role.
In one night we observed a spectrum of RW~Aur~A with exceptionally strong
shell and accretion components present in many lines of neutrals and ions,
which makes it possible to estimate the physical conditions
in different parts of the accretion stream.

In order to account for all the periodic 
phenomena, we
present two sketches of models, both dependent on the presence of
non--axisymmetric accretion.

The first model is based on the binary hypothesis, where the secondary is
assumed to generate enhanced accretion from one region of the disk,
from where the gas stream originates and finally hits the surface of the
primary. With new data on the inclination of the jet
(Dougados, private communication), which is presumably aligned with
the orbital axis, we get a solution where the
secondary is a brown dwarf moving at a distance of only 8--10 $R_{\sun}$
from the primary in a nearly circular orbit. The presence of this
secondary
is considered to be responsible for the exceptional properties
of RW~Aur compared to other CTTSs.

The second model assumes that RW~Aur~A is a single star with two
major accretion streams within a global magnetosphere that is tilted
relative to
the rotation axis or is intinsically non--axisymmetric.

Both these model concepts can explain quantitatively and qualitativly
most of the complex variations, but both also have their short-comings.
These
ideas are intended to provide a basis for further developments
of the theoretical concepts, and also for selecting key observational tests.

\begin{acknowledgements}

Thanks go to E.A.\ Kolotilov who kindly provided the UBV magnitudes
of comparison stars used in our CCD photometry of RW~Aur, to David Kennedal
who made the UBV photometry at the Swedish 60cm telescope, and to
P.\ Artymowicz and P.\ Carlqvist for
enlightening discussions. We also thank Catherine Dougados for providing
information on the jet prior to publication.
This work was supported by the Academy of
Finland, the Finnish Graduate School of Astronomy and Space Physics,
the Crafoord Foundation, the Swedish
Natural Science Foundation and the Portuguese foundation for Science
and Technology.

\end{acknowledgements}

\end{document}